\documentclass[preprint,pre,a4paper,superscriptaddress,nofootinbib]{revtex4-1}

\usepackage{graphicx}
\usepackage{amsmath}
\usepackage{latexsym}
\usepackage{amsfonts}
\usepackage{psfrag}
\usepackage{float}

\newcommand{\titlefont}{\fontfamily{ptm}\fontencoding{T1}\fontshape{n}\fontseries{b}\fontsize{18pt}{20pt}\selectfont}
\newcommand{\authorfont}{\fontfamily{ptm}\fontencoding{T1}\fontshape{n}\fontseries{b}\fontsize{14pt}{16pt}\selectfont}
\newcommand{\authoringfont}{\fontfamily{ptm}\fontencoding{T1}\fontshape{sl}\fontseries{m}\fontsize{12pt}{14pt}\selectfont}
\newcommand{\imtextfont}{\fontfamily{ptm}\fontencoding{T1}\fontshape{n}\fontseries{m}\fontsize{12pt}{16pt}\selectfont}

\newcommand{\indfont}{\fontfamily{ptm}\fontencoding{T1}\fontshape{n}\fontseries{m}\fontsize{8pt}{8pt}\selectfont}
\newcommand{\BEQ}{\begin{equation}}     % Gleichungen Anfang ..
\newcommand{\BEA}{\begin{eqnarray}}
\newcommand{\EEQ}{\end{equation}}       % .. und Ende
\newcommand{\EEA}{\end{eqnarray}}

\begin{document}

\imtextfont
\begin{center}
\titlefont{Numerical study of a slip-link model for polymer melts\\}
\vspace*{12pt}
\authorfont{Diego Del Biondo$^{\indfont \ddagger \dagger}$, Elian M. Masnada$^{\indfont \star}$,  Samy Merabia$^{\indfont \ddagger}$\footnote{E-mail: samy.merabia@univ-lyon1.fr}, Marc Couty$^{\indfont \dagger}$, Jean-Louis Barrat$^{\indfont \star,}$\footnote{E-mail: jean-louis.barrat@ujf-grenoble.fr}\\}
\vspace*{6pt}
\authoringfont{
$^{\indfont \ddagger}$ Universit\'e de Lyon; Univ. Lyon I,  Laboratoire de Physique de la Mati\`ere Condens\'ee et des Nanostructures; CNRS, UMR 5586, 43 Bvd. du 11 Nov. 1918, 69622 Villeurbanne Cedex, France\\ 
$^{\indfont \star}$ Universit\'e Grenoble I/CNRS, LIPhy UMR 5588, 38041 Grenoble, France\\
$^{\indfont \dagger}$ MFP MICHELIN 23, Place des Carmes-D\'echaux 63040 Clermont-Ferrand Cedex 9, France\\}
\end{center}

\date{\today}

\begin{abstract}
We present a numerical study of the slip link model introduced by Likhtman for describing the dynamics of dense polymer melts. After reviewing the technical aspects 
associated with the implementation of the model, we extend previous work in several directions. The dependence of the relaxation modulus with the slip link density and the 
slip link stiffness is reported. Then the nonlinear rheological properties of the model, for a particular set of parameters, are explored. Finally, we introduce 
excluded volume interactions in a mean field like manner in order to describe inhomogeneous systems, and we apply this description to a 
 simple  nanocomposite model.  
%For that, we introduce excluded volume interactions in a mean field like manner. We study the viscosity of a nanocomposite as function of the filler volume fraction. We consider bare fillers distributed on a regular ({\it i.e.}, cubic lattice) lattice. 
%The filler-monomer interaction is modeled by a repulsive potential. 
With this extension, the slip link model appears as a simple and generic model of a polymer melt, that 
can be used as an alternative to molecular dynamics for coarse grained simulations of complex polymeric systems.       
\\
\end{abstract}

\maketitle

\section{Introduction}

The mathematical description of rheological properties of entangled polymers is a difficult challenge, which can be addressed at several different levels
of accuracy and complexity.  The most popular  and very successful approach is  the one based on the so called tube model of Doi, Edwards and de Gennes~\cite{DoiEdwards1986}.
 In this model, the description is reduced to the motion of a single chain reptating along a tube that represents the topological constraints imposed by other chains. The 
model is partly analytic, introduces only a few parameters, and through some approximations can be converted into a local constitutive equation \cite{DoiEdwards1986}. Its 
drawbacks are its intrinsically mean field character (tube length fluctuations or constraint release are not considered) and the difficulty in extending it to various chain 
architectures.  The first aspect can be corrected in part, and further modifications of the model including tube length fluctuations and constraint 
release~\cite{Likhtman2002} achieve  quantitative agreement with the rheological   data for linear homopolymer melts, with additional parameters. The corresponding model 
stays at the one chain level, and developments such as extension to large strain rate, polydisperse or spatially inhomogeneous systems are difficult.  
  At the other extreme, a fully realistic modeling of the dynamical properties of a polymer melt can, in principle, be achieved using molecular dynamics or kinetic 
Monte-Carlo simulations of coarse grained polymer models~\cite{Kremer1990,Lahmar2009}. Such an approach has indeed provided numerical evidence for the validity of the 
reptation mechanism, and the analysis of the configurations allows one to identify the tube structure and the entanglements~\cite{Everaers2004}. However, the approach is so
 costly from a computational standpoint that the detailed study of rheological behavior is difficult, especially if one is interested in  deformation rates that are not 
large compared to the inverse reptation time of a chain.

Models intermediate between the tube description and the fully atomistic simulation have been proposed by several groups, and are generically described as ''slip link'' 
models. Such models inherit the tube model in the sense that they impose artificially the existence of topological constraints onto chain 
motion~\cite{Edwards1986,Rubinstein2002}. These topological constraints are, however, treated as statistical fluctuating objects that interact with the polymer chains, 
without modifying their equilibrium statistics. The polymer chains themselves are usually described as Rouse chains of Brownian particles connected by Hookean springs, and  
submitted to friction and random forces. Such models have the interest of easily accommodating complications such as polydispersity, complex chain architectures. They can 
also incorporate in a natural way constraint release and fluctuations in the tube length (or number of constraint per chain).  While originally intended to work as single 
chain models, they can also incorporate interchain interactions, as demonstrated below. They therefore offer an interesting compromise that preserves the computational 
simplicity of  tube models, but can be related more directly to an atomistic picture of the system.

Several different implementations of slip link models have been described in the literature, starting with the work of Marrucci and 
coworkers~\cite{Masubuchi2001,Oberdisse2002,Masubuchi2003,Yaoita2008}, Doi and Takimoto~\cite{DoiTakimoto2003} and 
including the model of Schieber and coworkers \cite{Schieber2003,Nair2006,Schieber2007} and of Likhtman \cite{Likhtman2005}. Here we concentrate on Likhtman's model, which
 we found to be particularly simple in its implementation and most easily extended to interacting chains. While Schieber and coworkers have published an extensive study of 
the flow properties in their model \cite{Schieber2007}, Likhtman's original work concentrated on equilibrium properties and allowed him to specify the values of the model 
parameters appropriate for the description of several polymers. The present study aims at extending Likhtman's work into several directions. First, we will investigate how 
the various parameters in the model affect the linear rheological properties. Then, we briefly investigate the nonlinear rheological properties of the model. 
Finally, we extend the previous model in order to study an inhomogeneous system, namely a filled entangled system. For that, we introduce interchain interactions via a simple density dependent interaction, as described in  \cite{Muller2008}. In that part, we consider bare fillers distributed on a cubic lattice and the effect of the fillers volume fraction on the viscosity is investigated. Before we discuss our results, the next section describes in some detail our implementation of the model.

\section{Numerical implementation of Likhtman's model}

\begin{figure}[H]
\centerline{\includegraphics*[width=8cm]{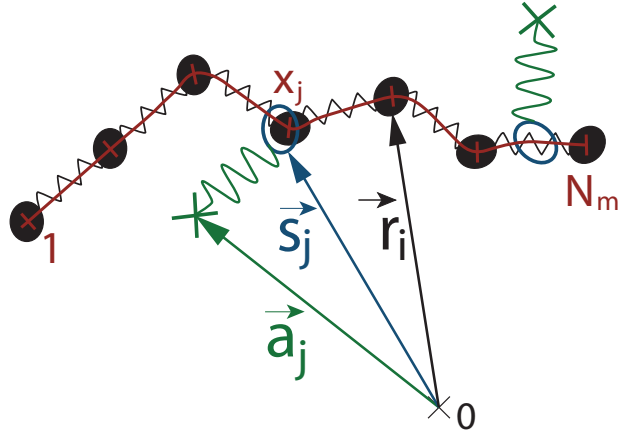}
\hspace{0.5cm}}
\caption{\label{chaine_SL} {\it {\footnotesize Rouse chain with slip-links. The ring of one slip-link is located by his curvilinear abscissa $x_j$. From $x_j$ the vector $\vec{s}_j$ is constructed according to equation \eqref{sj}. The anchoring points $\vec{a}_j$ are distributed around $\vec{s}_j$ with the following Boltzmann weight : $\exp{ \left( \frac{3k_BT}{2N_sb^2}(\vec{s}_j-\vec{a}_j)^2\right)}$}}}
\end{figure}

The original model of Likhtman involves an ensemble of noninteracting  Rouse chains, which are constrained by additional springs representing the topological constraints and
  called slip-links, as shown schematically in figure \ref{chaine_SL}. Each slip link is defined by a fixed  anchoring point at position $\vec {a}_j$ and a ring attached to 
the chain at position $\vec {s}_j$. The ring is constrained to move along the polymer chain by traveling along straight lines between adjacent monomers: 
\BEA
\label{sj}
	\vec{s}_j=\vec{r}_{{\rm trunc}(x_j)}+(x_j-{\rm trunc}(x_j)) (\vec{r}_{{\rm trunc}(x_j)+1}-\vec{r}_{{\rm trunc}(x_j)})
\EEA ${\rm trunc}(x)$ is the largest integer not greater than $x$ and $x_j$ is the curvilinear abscissa of the ring along its chain.\\
The anchoring point $\vec {a}_j$ are  fixed in space as long as the slip link $j$ is not destroyed.
Different destruction/creation rules for the slip links will be considered in this article and detailed later on. 
Each ring is connected to its anchoring point by a Hookean spring corresponding to the  confining potential $U_{SL}(\{\vec{s}_j\})=\frac{3k_BT}{2N_sb^2} (\vec{a}_j-\vec{s}_j)^2$, where 
${3k_BT}/{N_sb^2}$ is the slip slink stiffness, here counted in number of monomers $N_s$, $b$ being the monomer segment length and $k_BT$ the thermal energy.
The total potential felt by the single chain with $Z$ slip links writes then :
\BEA
	U = U_{\rm ROUSE} + U_{\rm SL}
\EEA
\BEA
\label{pot_rouse}
	U_{\rm ROUSE}(\{\vec{r}_i\})=\frac{3k_BT}{2\,b^2} \sum_{i=1}^{N_m} (\vec{r}_{i}-\vec{r}_{i-1})^2
\EEA
\BEA
\label{pot_SL}
	U_{\rm SL}(\{\vec{s}_j\})=\frac{3k_BT}{2N_sb^2} \sum_{j=1}^{Z} (\vec{a}_j-\vec{s}_j)^2
\EEA
where we note that the parabolic form of the slip link potential does not perturb the Gaussian statistics of the single chain. Given the total potential, the motion of 
monomer $i$ of the single chain obeys the Langevin equation: 
\BEA
\label{Langevin_mono}
	\xi \frac{d\vec{r}_i}{dt}=\frac{3k_BT}{b^2}(\vec{r}_{i+1}-2\vec{r}_i+\vec{r}_{i-1})+ \vec{\nabla}_{\vec{r}_i}U_{\rm SL} +\vec{f}_i(t)
\EEA
\BEA
	\vec{\nabla}_{\vec{r}_i}U_{\rm SL}=\frac{3k_BT}{N_sb^2}\sum_{j:{\rm trunc}(x_j)=i}(1-(x_j-{\rm trunc}(x_j)))(\vec{a}_j-\vec{s}_j) \nonumber\\ +\frac{3k_BT}{N_sb^2}\sum_{j:{\rm trunc}(x_j)=i-1}(x_j-{\rm trunc}(x_j))(\vec{a}_j-\vec{s}_j)
\EEA
\BEA
	<\vec{f}_i(t)>&=&\vec{0}
\EEA
\BEA
	<\vec{f}_{i}(t)\vec{f}_{j}(t')>&=&2\,\xi\,k_B\,T\,\delta_{ij}\delta(t-t') \mbox{I}
\EEA
where $\xi$ is the monomer friction coefficient. The force $\vec{\nabla}_{\vec{r}_i}U_{SL}$  on monomer $i$ is due to the slip links between the monomers 
$i-1$ and $i+1$. Finally, $\vec f_i$ is a random Brownian force, satisfying the usual fluctuation/dissipation relations ( $\mbox{I}$ denotes the unit tensor). The slip links
 positions obey a Langevin equation coupled to the previous one:
\BEA
\label{Langevin_SL}
	\xi_s \frac{dx_j}{dt}=-\vec{\nabla}_{x_j}U_{\rm SL}+g_j(t)
\EEA
\BEA
	-\vec{\nabla}_{x_j}U_{\rm SL}=\frac{3k_BT}{N_sb^2}(\vec{r}_{{\rm trunc}(x_j)+1}-\vec{r}_{{\rm trunc}(x_j)})(\vec{a}_j-\vec{s}_j)
\EEA
\BEA
	<g_i(t)>&=&0
\EEA
\BEA
	<g_{i}(t)g_{j}(t')>&=&2\,\xi_S\,k_B\,T\,\delta_{ij}\delta(t-t')
\EEA
where we have introduced the slip link friction coefficient $\xi_s$. The value of $\xi_s$ is chosen to be much smaller than the monomeric friction $\xi$, so that the 
diffusion of the slip links does not introduce a significant additional dissipation. The different parameters of the slip link model are summarized in the table 
\ref{table_parameters} below, and in the following we will study essentially how the two parameters $N_e$ and $N_s$ affect the
linear rheology of the model.
\newline

To close the presentation of the model, we present now the slip links renewal algorithm.  A static binary correspondence between pair slip links. When a slip link passes through the end of its chain, it is destroyed and instantaneously recreated at the extremity of a randomly chosen chain, where the extremity  is defined as the $N_e$ end monomers. Simultaneously, its companion is destroyed
 and instantaneously recreated at a random position of a randomly chosen chain. This renewal mechanism ensures that the slip links density is uniform along a chain, so that the   Gaussian statistics of the chains remains unaffected. 
A non uniform distribution of slip links would create stresses localized at the center of the polymer chain, which in turn would yield to the 
collapse of the chain compared to the initial Gaussian configuration. In addition, in a real polymer melt, the entanglements are not static but they can disappear and be created on time scales comparable to the 
chain relaxation times. These relaxation mechanisms, called constraint release (CR), are particularly important to explain
 the rheology of entangled polymer melts~\cite{Liu2006}. The static binary correspondence allows one to  account for this process without exaggerating it, as would be the case for a non-static binary correspondence. This renewal scheme has been validated by Likhtman by comparison with data for polystyrene~\cite{Likhtman2005}.

\begin{table}[H]

\begin{center}
\begin{tabular}{|c|c|}
\hline
temperature & $k_B\,T=1$ \\ \hline
monomer size & $b=1$ \\ \hline
friction coefficient of the entropic springs & $\xi=1$ \\ \hline
friction coefficient of the slip-links & $\xi_S=0.1\,\xi$ \\ \hline
characteristic time & $\tau_0=\frac{\xi\,b^2}{3\,\pi^2\,k_B\,T}$ \\ \hline
number of slip-links per chain & $Z=\frac{N_m}{N_e}$ \\ \hline
number of Kuhn's segments between slip-links & $N_e$ \\ \hline
stiffness of the slip-links & $\frac{3k_BT}{N_sb^2}$ \\ \hline
\end{tabular}
\end{center}
\label{table_parameters}
\caption{Main parameters that define the slip link model.}
\end{table}

\section{Influence of the model parameters on the relaxation modulus}

\begin{figure}[H]
\centerline{\includegraphics*[width=8cm]{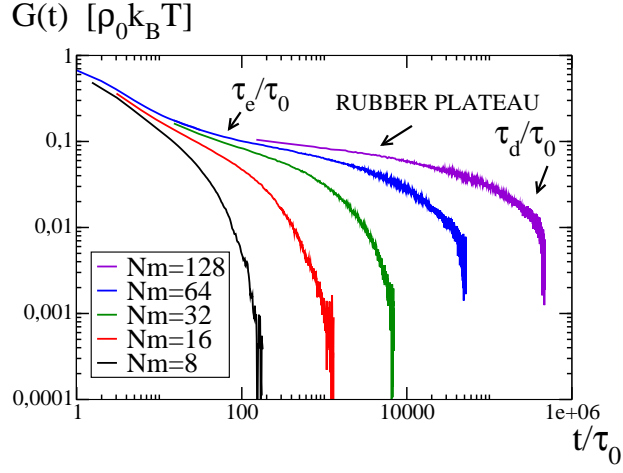}
\hspace{0.5cm}}
\caption{\label{G_Nm_varie} {\it {\footnotesize Stress relaxation modulus as a function of time for different chain lengths $N_m$. $N_e=4$ and $N_s=0.5$.}}}
\end{figure}

\begin{figure}[H]
\centerline{\includegraphics*[width=8cm]{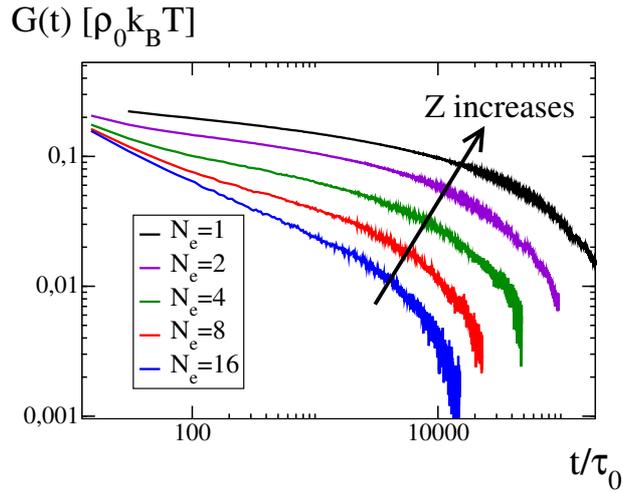}
\hspace{0.5cm}}
\caption{\label{G_Ne_varie} {\it {\footnotesize Stress relaxation modulus as a function of time for different values of the mean number of monomers between slip links $N_e$. Other parameters are $N_m=64$ and $N_s=0.5$.}}}
\end{figure}

\begin{figure}[H]
\centerline{\includegraphics*[width=8cm]{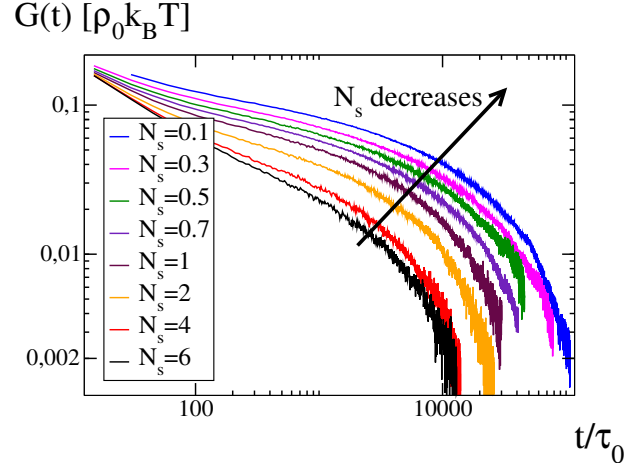}
\hspace{0.5cm}}
\caption{\label{G_Ns_varie} {\it {\footnotesize Stress relaxation modulus as a function of time for different slip link stiffness $N_s$. $N_m=64$ and $N_e=4$.}}}
\end{figure}

We now analyze the influence of the model parameters on the linear rheology. We focus on parameters $N_e$ and $N_s$ which control the density of effective entanglements in the slip link model. Following~\cite{Ramirez2007}, the expression of the shear relaxation modulus is given by :

\BEA
\label{modulebis}
        G(t)=\,\frac{V}{k_BT}\,\frac{1}{3}<\sum_{\alpha=1}^{2}\sum_{\beta>\alpha}^{3}\sigma_{\alpha\beta}^{\rm Rouse}(t)\,\sigma_{\alpha\beta}^{\rm T}(0)>
\EEA
where $\sigma_{\alpha\beta}^{\rm T}$ is the instantaneous shear stress defined by
\BEA
\label{total_stress}
\sigma_{\alpha\beta}^{\rm T}=\sigma_{\alpha\beta}^{\rm Rouse}+\sigma_{\alpha\beta}^{\rm SL}
\EEA
where $\sigma_{\alpha\beta}^{\rm SL}$ and $\sigma_{\alpha\beta}^{\rm Rouse}$ are the instantaneous shear stresses related to the $Z=N_m/N_e$ slip-links on the considered chain and 
the Rouse potential respectively. They are given by
\BEA
\label{contrainte}
        \sigma_{\alpha\beta}^{\rm Rouse}(t)\,=\,-\frac{1}{V}\,\sum_i\,<\alpha_{i}\;{F\beta_{i}}^{\rm Rouse}>
\EEA
\BEA
\label{contraintebis}
        \sigma_{\alpha\beta}^{\rm SL}(t)\,=\,-\frac{3k_BT}{N_s b^2 V}\,\sum_{j=1}^{Z}\,<(s_{\alpha,j}-a_{\alpha,j})(s_{\beta,j}-a_{\beta,j})>
\EEA
$V$ is the volume occupied by the polymer melt, $\alpha_{i}$ is the $\alpha$ coordinate vector of $\vec r_i$ and ${F\beta_{i}}^{\rm Rouse}$ is the $\beta$ coordinate vector of the total Rouse force $\vec {F_i}^{\rm Rouse}$ felt by monomer $i$. 
Let us  first consider FIG. \ref{G_Nm_varie} which displays the time evolution of the shear relaxation modulus for polymer chains of different lengths, here counted in number of monomers $N_m$.
The initial value of the modulus is $\rho_0 k_BT$, where $\rho_0$ is the mean density of the polymer melt.
The relaxation of the shortest chains considered is typical of unentangled polymer chains, and is close to the shear relaxation modulus predicted by the Rouse model.
On increasing the chain length, the shear modulus deviates from the Rouse model predictions.
In particular, for the two largest values of $N_m$ considered, a rubbery plateau appears, reminiscent of the plateau observed in rheological experiments of entangled polymer melts.
Note, however the log-log scale of FIG. \ref{G_Nm_varie}, which implies that the rubbery plateau still involves a significant relaxation for $N_m=128$ .
FIG. \ref{G_Ne_varie} shows the evolution of the shear relaxation modulus when increasing the number of slip-links per chain $Z=N_m/N_e$.
When $Z$ is small, the effect of the slip links on the dynamics of the chain is weak, and the shear relaxation modulus is closer to the Rouse predictions, with the absence of a well defined plateau. On increasing $Z$, a 
rubbery plateau appears : its amplitude increases with $Z$ and simultaneously the
 final relaxation time $\tau_d$ increases (see FIG. \ref{G_Ne_varie}). Alternatively, the slip-links stiffness may be increased ($N_s$) at given $N_e$ (or Z), which also results in an
 increase of the modulus and of the terminal time (see FIG. \ref{G_Ns_varie}). Note however that for values of $N_s$ that stay ''physical'' ($N_s >1$) this increase is relatively modest.

To make the discussion more quantitative, we have extracted a terminal time $\tau_d$ and an amplitude $G_N^{(0)}$ from the simulated $G(t)$ using a fitting procedure with a simple tube model. The reptation model \cite{DoiEdwards1986} predicts the evolution of the relaxation modulus :\\
\BEA
\label{fitG1}
        G(t)=G_N^{(0)} \Psi(t)
\EEA
\BEA
\label{fitG2}
        \Psi(t) = \sum_{p\; odd}\frac{8}{p^2\pi^2}\exp\left(\frac{-p^2t}{\tau_d}\right)
\EEA
In the reptation model, the rubbery modulus $G_n^0$ and the terminal time $\tau_d$ are related to the distance $a$ between entanglements through:
$G_n^0 = \rho_0 k_BT b^2/a^2 \propto M_e^{-1}$ and $\tau_d =\frac{\zeta N^3 b^2}{\pi^2 k_BT} b^2/a^2 \propto M_e^{-1}$ where $M_e= a^2/b^2$ denotes the mean number of monomers between two entanglements and the other parameters $\rho_0$ and $\zeta$ denote respectively the monomer density and 
the monomeric friction. The topology of the tube is described by a single phenomenological parameter $a$, while in the slip link model, the effective tube depends on the two
 parameters $N_e$ and $N_s$. Therefore the correlation between $G_N^{(0)}$ and $\tau_d$ that exists in the reptation model is absent from the slip link model. In the 
following, we will simply use equations \ref{fitG1} and \ref{fitG2} as convenient fitting formulae, including in situations 
where the chains display a "quasi-Rouse" dynamics (for large $N_e$ for example).
In the other extreme case of well entangled polymer melts, $G_N^{(0)}$ can be interpreted as a plateau modulus and this fitting formulae leading to treat $G_N^{(0)}$  and $\tau_d$ as independent parameters.

\begin{figure}[H]
\centerline{\includegraphics*[width=8cm]{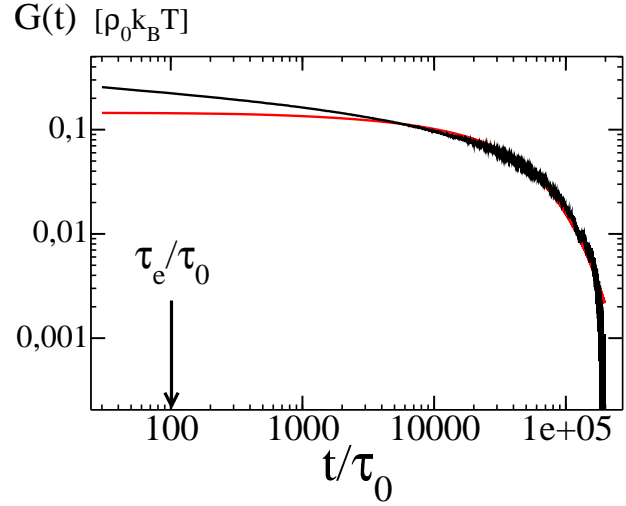}
\hspace{0.5cm}}
\caption{\label{fit} {\it {\footnotesize Fitting procedure to obtain the reptation parameters $G_N^{(0)}$ and $\tau_d$ from the stress relaxation modulus. The black curve is the simulated relaxation modulus for $N_m=64$, $N_e=1$ and $N_s=0.5$. The red curve is the best fit of G(t) using the reptation model eqs.~(\ref{fitG1}) and (\ref{fitG2}).}}}
\end{figure}

%%%%%%%%%%%%%%%%%%%%%%%%%%%%%%%%%%%%%%%%%%%%%%%%%%%%%%%%%%%%%%%%%%%%%%%%%%%%%%%%%%%%%%%%%%%%%%%%%%%%%%%%%%%%%%%%%%
%%%%%%%%%%%%%%%%%%%%%%%%%%%%%%%%%%%%%%%%%%%%%%%%%%%%%%%%%%%%%%%%%%%%%%%%%%%%%%%%%%%%%%%%%%%%%%%%%%%%%%%%%%%%%%%%%%
%%%%%%%%%%%%%%%%%%%%%%%%%%%%%%%%%%%%%%%%%%%%%%%%%%%%%%%%%%%%%%%%%%%%%%%%%%%%%%%%%%%%%%%%%%%%%%%%%%%%%%%%%%%%%%%%%%
The simple form  \ref{fitG2} is supposed to predict $G(t)$ for times longer than a time $\tau_E$, which in the reptation model is the Rouse time corresponding of a subchain 
of $M_e$ monomers. Fig~\ref{fit} compares the simulated $G(t)$ to  equation ~\eqref{fitG1} and eq.~\eqref{fitG2} where we have retained only the ten first terms in the sum 
and we have emphasized the time domain used in the fit $t>\tau_E$. The sum of exponentials describes well the simulated $G(t)$ for long times, while close to the lower boundary $t \simeq \tau_E$ the fit deviates from $G(t)$, due to short time Rouse like contributions.  Hence, we conclude that the exponential form eq.~\eqref{fitG1} is an 
acceptable fit of our simulation data provided there is a clear separation of time scales $\tau_E \ll \tau_d$. This will be the case for long chains, small $N_e$ and large slip link stiffness (small $N_s$). This fitting procedure has been applied systematically to all the previous curves to extract $G_N^0$ and $\tau_d$ as a function of $N_e$ 
and $N_s$ even for quasi-Rouse relaxation. 
%%%%%%%%%%%%%%%%%%%%%%%%%%%%%%%%%%%%%%%%%%%%%%%%%%%%%%%%%%%%%%%%%%%%%%%%%%%%%%%%%%%%%%%%%%%%%%%%%%%%%%%%%%%%%%%%%%
%%%%%%%%%%%%%%%%%%%%%%%%%%%%%%%%%%%%%%%%%%%%%%%%%%%%%%%%%%%%%%%%%%%%%%%%%%%%%%%%%%%%%%%%%%%%%%%%%%%%%%%%%%%%%%%%%%
%%%%%%%%%%%%%%%%%%%%%%%%%%%%%%%%%%%%%%%%%%%%%%%%%%%%%%%%%%%%%%%%%%%%%%%%%%%%%%%%%%%%%%%%%%%%%%%%%%%%%%%%%%%%%%%%%%

\begin{figure}[H]
\centerline{\includegraphics*[width=8cm,angle=-90]{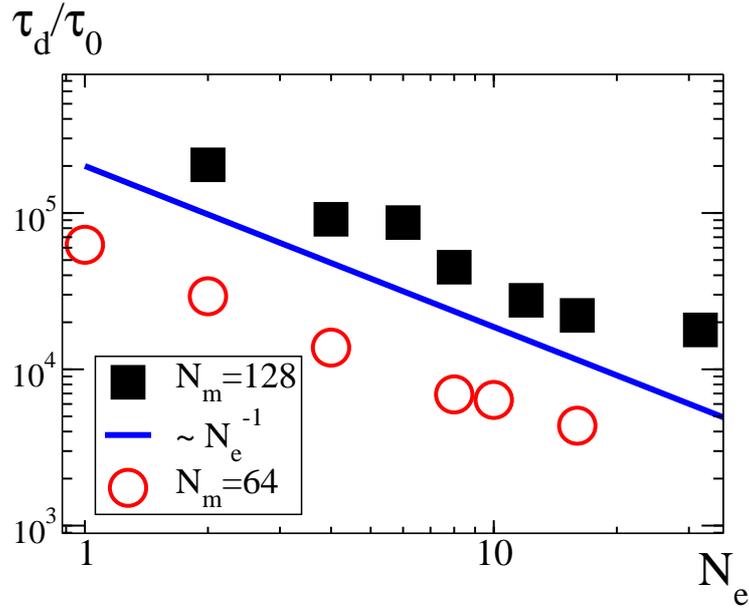}
\hspace{0.5cm}}
\caption{\label{taud_Ne_varie} {\it {\footnotesize Relaxation time $\tau_d/\tau_0$ as a function of $N_e$ for $N_s=0.5$. For $N_m=64$, we observe $\tau_d/\tau_0 \sim N_e^{-1.19}$ while for $N_m=128$, $\tau_d/\tau_0 \sim N_e^{-0.99}$.}}}
\end{figure}

\begin{figure}[H]
\centerline{\includegraphics*[width=8cm,angle=-90]{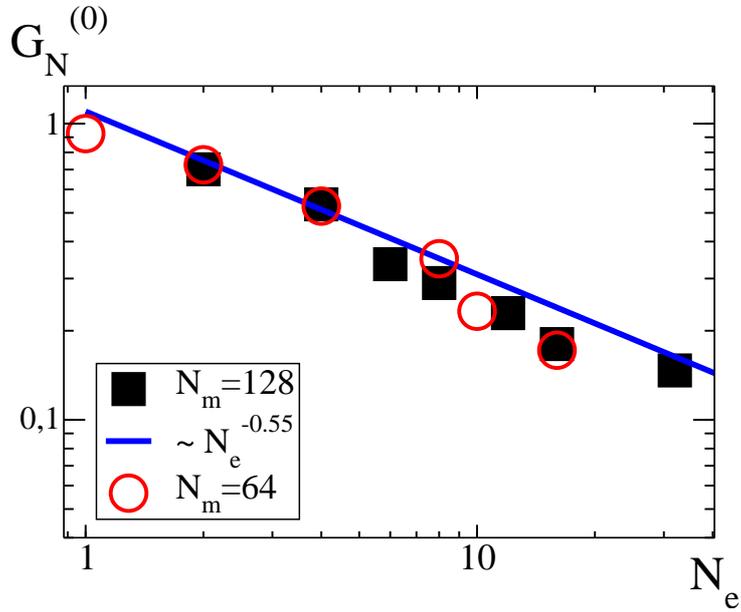}
\hspace{0.5cm}}
\caption{\label{Gn0_Ne_varie} {\it {\footnotesize Amplitude $G_N^{(0)}$ obtained with the fitting procedure illustrated in fig. 
\ref{fit} as a function of the parameter $N_e$, for two chain lengths: $N_m=64$ and $N_m=128$. The parameter $N_s=0.5$ is fixed. For $N_m=64$, we observe $G_N^{(0)} \sim N_e^{-0.56}$, while for $N_m=128$, $G_N^{(0)} \sim N_e^{-0.59}$.}}}
\end{figure}

\begin{figure}[H]
\centerline{\includegraphics*[width=8cm,angle=-90]{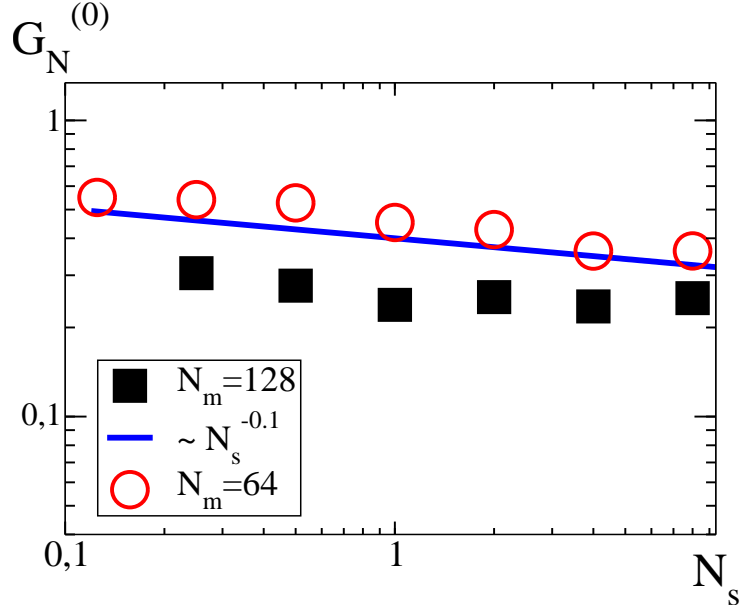}
\hspace{0.5cm}}
\caption{\label{Gn0_Ns_varie} {\it {\footnotesize Amplitude $G_N^{(0)}$ as function of the parameter $N_s$ and with $N_e=4$, for $N_m=64$, the power law obtained is $G_N^{(0)} \sim N_s^{-0.10}$. For $N_m=128$, we chose $N_e=8$ the power law is $G_N^{(0)} \sim N_s^{-0.06}$.}}}
\end{figure}

\begin{figure}[H]
\centerline{\includegraphics*[width=8cm,angle=-90]{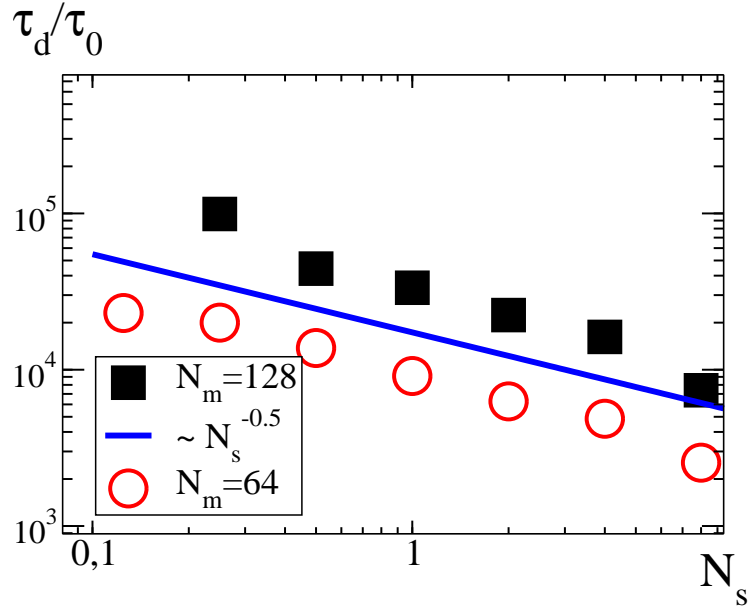}
\hspace{0.5cm}}
\caption{\label{taud_Ns_varie} {\it {\footnotesize Relaxation time $\tau_d/\tau_0$ as a function of $N_s$. For $N_m=64$, $N_e=4$ we observe $\tau_d/\tau_0 \sim N_s^{-0.52}$ while for $N_m=128$ and $N_e=8$, $\tau_d/\tau_0 \sim N_s^{-0.56}$.}}}
\end{figure}

The resulting values of $\tau_d$ and $G_N^{(0)}$ as function of $N_s$ and $N_e$ are reported in figures \ref{taud_Ne_varie},  \ref{Gn0_Ne_varie}, \ref{Gn0_Ns_varie}, and \ref{taud_Ns_varie} for two chain lengths $N_m=64$ and $N_m=128$.   We mention here  that  the dependence of the viscosity and diffusion coefficient  on molecular weight have been analyzed in reference \cite{Likhtman2005}, and shown to be in good agreement with the experimental trends, with a crossover from Rouse behavior to $\sim N_m^{3.6}$ behaviour for the viscosity. Our results  confirm this analysis, which is not reproduced here.

The dependence of $\tau_d$ on  $N_e$, reported in figure \ref{taud_Ne_varie}, is consistent with the general expectation from the  reptation model, namely  $\tau_d \sim N_e^{-1}$. This reflects the fact that, for the value 
of $N_s$ that is used, each slip link is acting as an independent topological constraint, through which 
the chain has to travel in order to disentangle.  The  length of the primitive tube depends linearly on $N_e$. On the other hand,  figure  \ref{Gn0_Ne_varie} shows that the amplitude $G_N^{(0)}$ has a much weaker dependence on $N_e$ than expected in the simple reptation picture, provided we have identified $G_N^0$ with the plateau modulus.

 When the slip links are dense along the chain, they do not act as independent crosslinks. Therefore identifying directly the slip links with entanglements is not possible, except perhaps in the asymptotic limit of large $N_m$ and $N_e\gg 1$, which is not explored here in view of the associated computational cost.
The amplitude $G_N^{(0)}$ depends weakly on the chain length at a fixed value of $N_e$ and $N_s$, as seen in figures \ref{Gn0_Ne_varie} and \ref{Gn0_Ns_varie}.  This is 
consistent with the reptation picture, where the rubbery plateau is entirely controlled by the density of slip links and their stiffness independently of the mass of the chain. 
The variation of $\tau_d$ with $N_s$ is reported in figure \ref{taud_Ns_varie}. For the two chain lengths studied, $\tau_d$ decreases algebraically with $N_s$ with an exponent close to $-0.6$.

In the following, we will concentrate on a particular set of parameters, $N_s=0.5$, $N_e=4$,  which was shown by Likhtman \cite{Likhtman2005} to give  an appropriate description of polystyrene data. Our study shows, however, 
that the slip link models offer a large flexibility that goes beyond that of tube models, with in particular the 
ability to vary independently the amplitude $G_N^{(0)}$ and the terminal time $\tau_d$, by playing with the independent parameters $N_e$ and $N_s$.

\section{Nonlinear rheological behavior}
In this section, we investigate  the non linear rheology of the slip link model.
% The parameters are those used for polystyrene melts in ref. \cite{Likhtman2005}.
To this end, we have applied a steady shear flow with a constant shear rate $\dot{\gamma}(t)=\dot{\gamma}$. The equations of motion obeyed by the 
monomers become: 
\BEA
        v_x=y \dot{\gamma}(t)+\frac{1}{\xi}\left({F}_x^{Rouse}+{F}_x^{SL}\right) \nonumber \\
        v_y=\frac{1}{\xi}\left({F}_y^{Rouse}+{F}_y^{SL}\right) \nonumber \\
        v_z=\frac{1}{\xi}\left({F}_z^{Rouse}+{F}_z^{SL}\right) \nonumber
\EEA
where we have added the term $\,y\,\dot{\gamma}(t)$ which convects the monomers in the imposed flow field. 
Simultaneously, the anchoring points of the slip-links are convected according to:
\BEA
        \vec{v}(X_{\vec{a}_j})=Y_{\vec{a}_j}\,\dot{\gamma}(t)\,\vec{e}_x \nonumber
\EEA
In addition, Lees-Edwards periodic boundary conditions are applied to all the monomers\cite{AllenTildesley1987}.\\

Finally, we consider here only the Rouse contribution to the instantaneous shear stress: $\sigma_{xy}=\sigma_{xy}^{\rm Rouse}$ and disregard the slip-links contribution $\sigma_{xy}^{\rm SL}$. Indeed, taking into account a total shear stress defined by $\sigma_{xy}^{\rm Rouse}+\sigma_{xy}^{\rm SL}$ does not quantitatively change the power laws characterizing the rheology of the polymer model, as it will be apparent later on. 

%In Fig.(\ref{histo_SL}), we give the slip-links distribution on a chain for two different shear rates namely 
%$\dot{\gamma}\tau_0=10^{-2}$ and $\dot{\gamma}\tau_0=10^{-5}$. At small $\dot{\gamma}\tau_0$, this distribution remains uniform while at larger $\dot{\gamma}\tau_0$ the density of slip-links at the chain extremity becomes higher. Indeed, because of the shear, the chains are stretched and tend to line up while the slip-links are 
%advected by the flow (see Fig.\ref{advicted}). As a consequence, the lifetime of the slip-links decreases when the shear rate 
%increases. In this way and with the slip link renewal rules described before, the model should take into account 
%the convective constraint release (CCR) mechanism proposed by Marrucci~\cite{Marrucci1996,Mead1998} and which describes the non-linear rheology of entangled polymers. 

\begin{figure}[H]
\centerline{\includegraphics*[width=8cm]{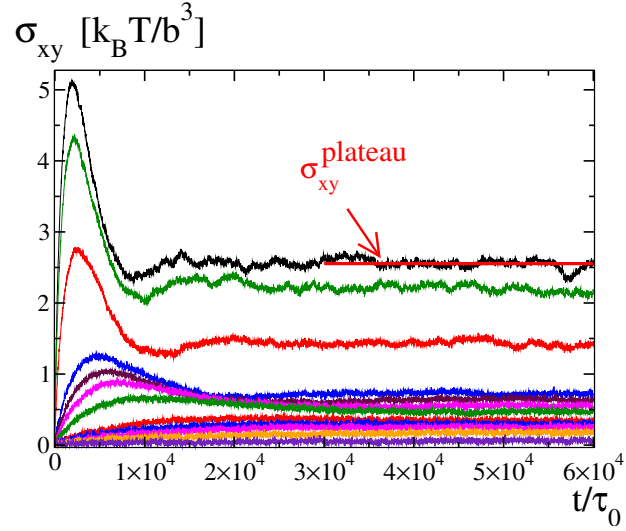}
\hspace{0.5cm}}
\caption{\label{sigma_xy_64_rouse} {\it {\footnotesize Shear stress as a function of time under steady shear flow at several shear rates. The mparameters are: $N_m=64$, $N_e=4$ and $N_s=0.5$. From top to bottom, the shear rates are $\dot{\gamma} \tau_0 = 10^{-2},8 \; 10^{-3},4 \;10^{-3},10^{-3},7\;10^{-4},5\;10^{-4},3\;10^{-4},10^{-4},7 \;10^{-5},5\;10^{-5},3\;10^{-5},10^{-5}$.}}}
\end{figure}

\begin{figure}[H]
\centerline{\includegraphics*[width=8cm]{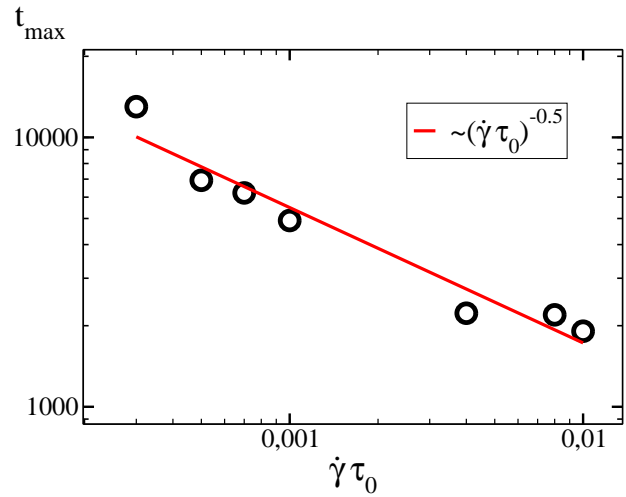}
\hspace{0.5cm}}
\caption{\label{timeovershoot} {\it {\footnotesize Time corresponding to the maximum of the stress overshoot (see FIG. \ref{sigma_xy_64_rouse}) as a function of the shear rate. The exponent is not sensitive to the definition of the shear stress ($\sigma_{xy}^{\rm Rouse}$ or $\sigma_{xy}^{\rm Rouse}+\sigma_{xy}^{\rm SL}$). Parameters~: $N_m=64$, $N_e=4$, $N_s=0.5$.}}}
\end{figure}

\begin{figure}[H]
\centerline{\includegraphics*[width=8cm]{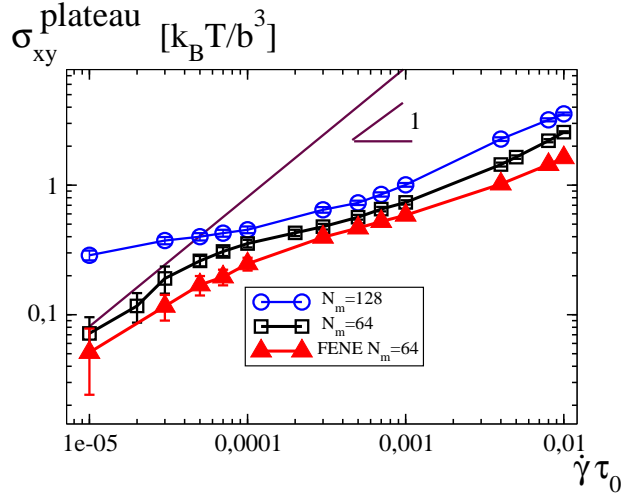}
\hspace{0.5cm}}
\caption{\label{log_stress_rate_bin} {\it {\footnotesize Evolution of the shear plateau as a function of  shear rate for Rouse chains of lengths 
$N_m=128$ ($\circ$); $N_m=64$ ($\square$) and FENE chains having length $N_m=64$ ($\triangle$). Solid lines are guides to the eye. 
The other parameters are: $N_e=4$ and $N_s=0.5$ }}}
\end{figure}

\begin{figure}[H]
\centerline{\includegraphics*[width=8cm]{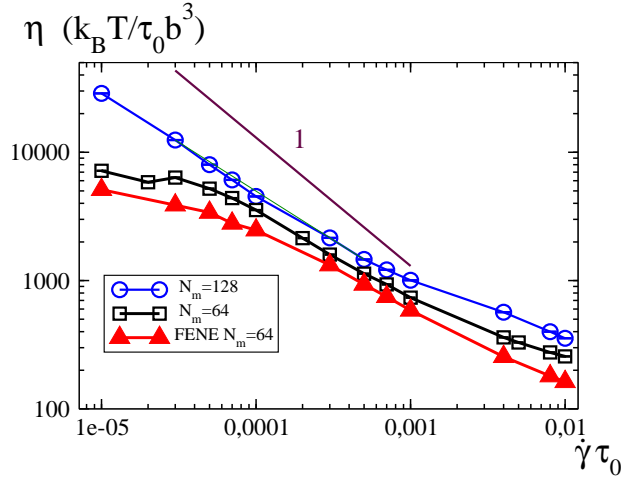}
\hspace{0.5cm}}
\caption{\label{viscosity_rate} {\it {\footnotesize Viscosity extracted from fig.~\ref{log_stress_rate_bin} for Rouse chains of lengths $N_m=128$ ($\circ$); $N_m=64$ ($\square$) and FENE chains having length $N_m=64$ ($\triangle$). The prediction of the convective constraint release model  of Marrucci~\cite{Marrucci1996},  $\eta \sim \dot{\gamma}$ is also shown. Our results correspond to $\eta \sim \dot{\gamma}^{-0.67}$ with $N_m=64$. 
All the results have been obtained using the Rouse expression of the shear stress $\sigma_{xy}^{\rm Rouse}$.
The extra contribution of the slip-links $\sigma_{xy}^{\rm SL}$ to the shear stress changes the shear thinning exponent from $0.66$ to $0.67$, in the simulations with $N_m=64$.
Same parameters as fig.\ref{log_stress_rate_bin}.}}}
\end{figure}

In figure~\ref{sigma_xy_64_rouse}, we have reported the evolution of the shear stress $\sigma_{xy}$ as a function of time under steady shear flow at several shear rates. The values of the shear rates considered range from $\dot{\gamma} \sim \tau_E^{-1}$ to $\dot{\gamma} \sim \tau_d^{-1}$ for which the chain has totally relaxed. 
Two situations have to be distinguished depending on the value of the shear rate $\dot{\gamma}$. 
For the largest shear rates, the evolution of the shear stress with time (or shear strain) is non-monotonous : the stress increases up to a maximum (the so called stress overshoot maximum). Then, the shear stress decreases to finally reach a plateau, which corresponds to a steady state situation. 
The existence of a stress overshoot for entangled polymers is well known experimentally and also predicted in the theoretical analysis of
Doi and Edwards who have considered the affine deformation of the (primitive chain) tube created by the entanglements in a shear flow and by the convective constraint release CCR model of Marrucci~\cite{Marrucci1996,Mead1998}. 
In the Doi-Edwards model, the stress overshoot occurs at a constant deformation $\gamma_{\max}=\dot{\gamma} t_{\rm max}\simeq 2$ and thus the 
time corresponding to the stress maximum scales as $t_{\rm max} \propto \dot{\gamma}^{-1}$. We have compared this prediction to our simulations in figure \ref{timeovershoot} where we observe that $t_{\rm max} \propto \dot{\gamma}^{-\kappa}$ with $\kappa \simeq 0.5$, meaning that the deformation at the overshoot increases with strain rate. This increase is also observed in experiments~\cite{MenezesGraessley1982,Schweizer2004} and in the Marrucci model at large strain rates. However, note that the scaling $t_{\rm max} \sim \dot{\gamma}$ is observed experimentally for extremely small shear rates $\dot{\gamma} \tau_d \ll 1$, a regime difficult to attain in our model. Note also that the value of the exponent $\kappa$ does not change if we include the contribution of the slip-links in the definition of the instantaneous shear stress.
Coming back to figure ~\ref{sigma_xy_64_rouse}, we observe at low shear rates the absence of stress overshoot, and a monotonous  
evolution of the shear stress: the stress increases before reaching a low steady state shear stress.
The values of the shear stress plateau as function of the shear rate (flow curve)  are reported in figure  \ref{log_stress_rate_bin} for two chain lengths $N_m=64$ and $N_m=128$ and for a finite extensible non linear elastic FENE chain with slip-links. In the latter model, the hookean springs between monomers are replaced by a non linear spring force which derives from 
the potential $U_{\rm FENE}(r)=-\frac{3k_BT}{2b^2} R_0^2 \log(1-(r/R_0)^2)$ which defines the maximal extension of the springs $R_0$ (we have set $R_0=1.6b$).
The evolution of the steady shear stress as a function of the shear rate displays three regimes :
At low shear rates, the shear stress increases approximately linearly with the shear rate at least for the chains of length $N_m=64$.
In this regime, the chains have totally relaxed in the typical shear time scale and the rheology of the melt is Newtonian: $\sigma^{\rm{plateau}} = \eta \dot{\gamma}$, $\eta$ being the viscosity of the melt of chains. This regime is not seen for the longest chains $N_m =128$, as it would correspond to 
very low shear rates that would need very long simulation times. 
For intermediate shear rates $\dot{\gamma} \tau_0 \in [10^{-4}; 10^{-3}]$, the evolution of the shear stress with the shear rate is slower: we observe $\sigma^{\rm{plateau}} \propto 
\dot{\gamma}^{\alpha}$ with an exponent $\alpha \simeq 0.3$ independent of the chain length and independent of the type of elastic springs.
This contrast with the Doi Edwards model which predicts in this intermediate shear rates range a decrease of the stress with the shear rate, which would lead  to a flow instability that is not usually observed in polymer melts. Again, CCR mechanisms are thought to restore the monotonicity of the flow curve yielding an effective viscosity $\eta \sim \dot{\gamma}^{-1}$ at large shear rates~\cite{Marrucci1996}. We have considered in fig.~\ref{viscosity_rate} the evolution of the viscosity as a function of the shear rate in steady state conditions. It turns out that the SL model displays a shear thinning behaviour less marked than predicted by Marrucci: in particular, we observe $\eta \sim \dot{\gamma}^{-x}$ with $x \simeq 0.7$ for the $N_m=128$ melt. 
%A possible reason for this small discrepancy with the theory 
%may come from the fact that the polymer chains considered here are moderately entangled with $Z=32$ and it is known  that the apparent exponent characterizing the shear thinning behaviour of entangled polymer is very sensitive to the degree of entanglement~\cite{Larson1999}. 
Finally, for the highest $\dot{\gamma}$, the polymer chains have also a shear thinning behavior with an apparent exponent $x \simeq 0.6$ for all the polymer models considered. These shear-thinning exponents  can be compared with  rheological measurements, which conclude $x=0.85$ for a polymer melt with a comparable degree of entanglement $Z=15$ \cite{Schweizer2004}. 

The study of the steady state viscosity gives also the opportunity to quantify the influence of the stress due to the slip links on the shear-thinning behavior. We have observed that if we use the expression of the shear stress which includes the contribution of the slip-links: $\sigma_{xy}^{\rm Rouse}+\sigma_{xy}^{\rm SL}$, the shear thinning exponent changes mildly passing from $0.67$ to $0.68$. The absolute value of the viscosity obtained from the two contributions is larger than that obtained with $\sigma_{xy}^{Rouse}$, by around $20$\% for $\dot{\gamma}\tau_0=10^{-5}$ and $10$\% for $\dot{\gamma}\tau_0=10^{-2}$, $(\sigma_{xy}^{\rm Rouse}+\sigma_{xy}^{\rm SL})/\sigma_{xy}^{\rm Rouse}$ evolving as $(\dot{\gamma}\tau_0)^{-0.01}$.

\begin{figure}[H]
\centerline{\includegraphics*[width=8cm]{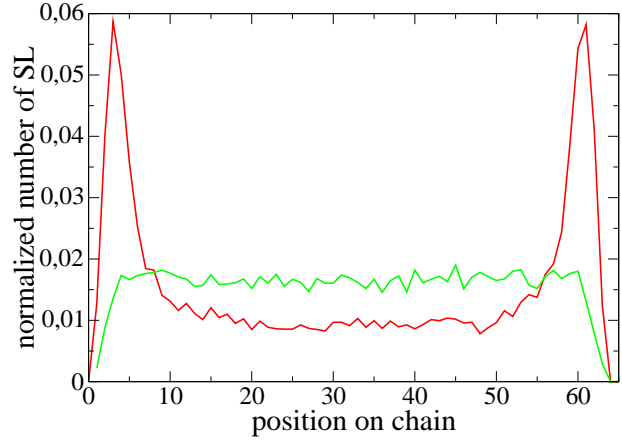}
\hspace{0.5cm}}
\caption{\label{histo_SL} {\it {\footnotesize Slip-link distribution along a chain for $\dot{\gamma}\tau_0=10^{-2}$ (red) and $\dot{\gamma}\tau_0=10^{-5}$
(green).  The distribution is uniform for low $\dot{\gamma}$ while it becomes non uniform under the strong shear flow. The model parameters are: $N_m=64$, $N_e=4$ and $N_s=0.5$. }}}
\end{figure}

\begin{figure}[H]
\centerline{\includegraphics*[width=8cm]{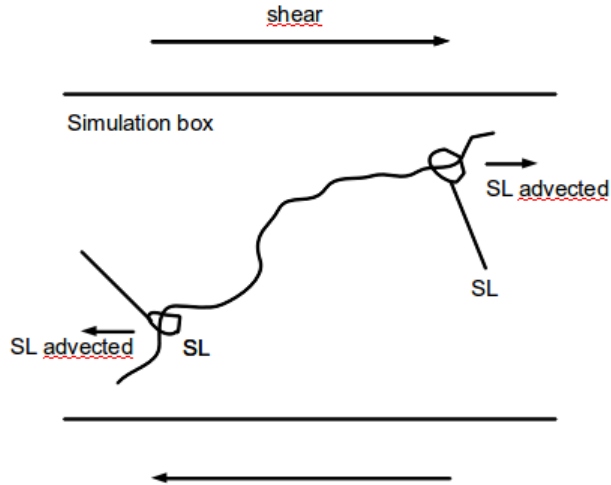}
\hspace{0.5cm}}
\caption{\label{advicted} {\it {\footnotesize Typical configuration of a chain (in the frame of the center of mass)  under strong shear flow conditions. As the slip-links are advected by the flow, they tend to accumulate at the chain extremities, which explains the non-uniformity observed for large shear rates (see Fig.\ref{histo_SL}).}}}
\end{figure}

\begin{figure}[H]
\centerline{\includegraphics*[width=8cm]{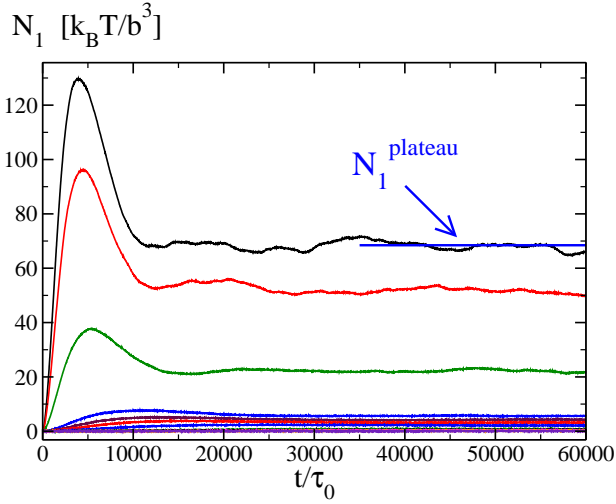}
\hspace{0.5cm}}
\caption{\label{N1_64} {\it {\footnotesize Evolution of the first normal stress difference ($N_1=\sigma_{xx}^{\rm Rouse}-\sigma_{yy}^{\rm Rouse}$) as a function of time. Parameters : $N_m=64$, $N_e=4$, $N_s=0.5$. From top to bottom, the shear rates are equal to $\dot{\gamma} \tau_0 = 10^{-2},8\;10^{-3},4\;10^{-3},10^{-3},7\;10^{-4},5\;10^{-4},3\;10^{-4},10^{-4},7\;10^{-5},5\;10^{-5},3 \;10^{-5},10^{-5}$. 
}}}
\end{figure}

\begin{figure}[H]
\centerline{\includegraphics*[width=8cm]{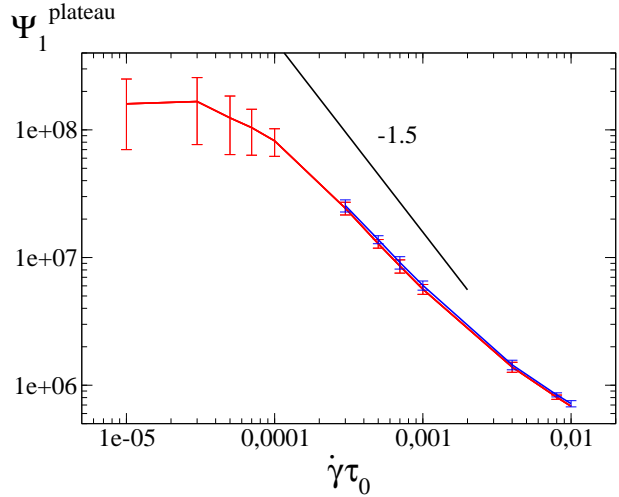}
\hspace{0.5cm}}
\caption{\label{PSI1_64} {\it {\footnotesize Evolution of the first normal stress coefficient plateau ($\Psi_1^{plateau}={N_1}/{(\dot{\gamma}\tau_0)^2}$) as a function of the
 shear rate. In red, the shear stress is given by $\sigma_{xy}^{\rm Rouse}$ while in blue the definition is $\sigma_{xy}^{\rm Rouse}+\sigma_{xy}^{\rm SL}$. We have also shown the 
theoretical scaling predicted by Marrucci~\cite{Marrucci1996}: $\Psi_1 \sim \dot{\gamma}^{-1.5}$. In our simulations we obtain $\Psi_1 \sim \dot{\gamma}^{-1.2}$. This exponent 
does not change with the definition of the shear stress. Parameters are: $N_m=64$, $N_e=4$, $N_s=0.5$. 
}}}
\end{figure}

 Finally, it is also important to stress at this point that, depending on the flow strength $\dot{\gamma}$ the steady slip-links distribution on the chain may become non-uniform. 
At small $\dot{\gamma}$, figure (\ref{histo_SL}) clearly shows that the slip-links are uniformly distributed along the chains as in equilibrium simulations. 
On the other hand, at larger $\dot{\gamma}$ slip-links tend to accumulate close to the chain extremities, while a depletion is observed at the centers, as seen in Fig.(\ref{histo_SL}). This may be understood as follows: Under strong shear flow the polymer chains are stretched and tend  to align with the stream lines, while the slip-links anchoring points are advected affinely by the flow (see Fig.(\ref{advicted})). As a consequence, the slip links tend to drift to the chain ends, and their lifetime of the slip-links decreases when the shear rate increases.
%{\bf je pense il faut dire un peu pleu ici which ones ? SL in the center ?}.\\ 
Apart from shear thinning, the non linear rheology of entangled polymer melts is accompanied by the development of normal stresses. This is quantified by the first and second normal stresses defined by:
\begin{equation}
N_1 = \sigma_{xx}^{\rm Rouse}-\sigma_{yy}^{\rm Rouse}
\end{equation}
and 
\begin{equation}
N_2 = \sigma_{yy}^{\rm Rouse}-\sigma_{zz}^{\rm Rouse}
\end{equation} 
or by the corresponding  first and second normal coefficients:
\begin{equation}
\psi_{1,2}(\dot{\gamma}) = N_{1,2}(\dot{\gamma})/\dot{\gamma}^2
\end{equation}
%and 
%\begin{equation}
%\psi_2(\dot{\gamma}) = N_2(\dot{\gamma})/\dot{\gamma}^2
%\end{equation}
In these  equations, $N_1(\dot{\gamma})$ and $N_2(\dot{\gamma})$ denote the steady state values of the normal stresses at a given shear rate. 
We have measured the normal stresses during shear start flow in fig.\ref{N1_64} for the same range of shear rates considered before. For high shear rates, 
the evolution of $N_1$ is non monotonous. The first normal stress difference increases before reaching a maximum which is observed after the stress overshoot maximum,
 the corresponding time $t'_{\rm max}$ being found to be nearly independent of the shear rate in agreement with the CCR model~\cite{Schweizer2004}. After this overshoot, the normal stress $N_1$ decreases to reach a steady state value 
$N_1(\dot{\gamma})$ which increases with the shear rate. Note that for small shear rates, the evolution of $N_1$ towards its steady state value is monotonous. 
The shear rate dependence of the plateau value of $N_1$ is best quantified by the normal stress coefficient $\psi_1$ defined above and calculated in fig.~\ref{PSI1_64}. 
At low shear rates, $\psi_1$ is approximately constant as expected for the reptation model when $\dot{\gamma} \tau_d \simeq 1$. For stronger shear flows, $\psi_1$ decreases with the shear rate $\dot{\gamma}$. For the sake of comparison, we have plotted in fig.\ref{PSI1_64} the scaling law $\psi_1 \sim \dot{\gamma}^{-1}$ predicted by the CCR model and observed experimentally~\cite{MenezesGraessley1980}. The simulation values of $\psi_1$ are in reasonable agreement with this scaling law at intermediate shear rates. 
Again the  disagreement at higher strain rates between the SL model results and the expected behaviour may be due to the relative small separation of time scales in our model between the reptation time $\tau_d$ and the Rouse time corresponding to the distance between slip links $\tau_E \sim 100\tau_0$.

\begin{figure}[H]
\centerline{\includegraphics*[width=8cm]{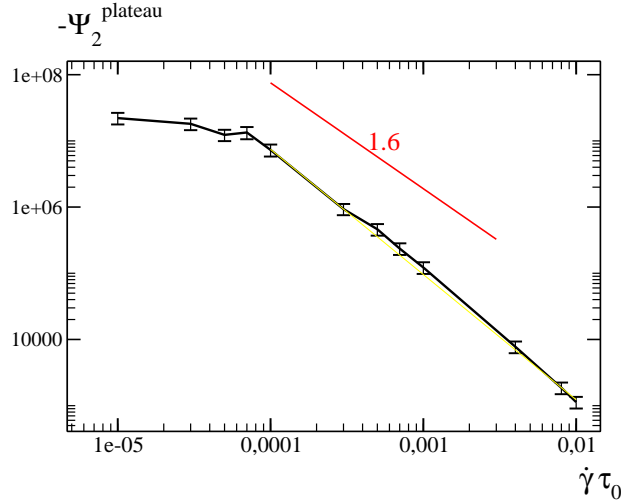}
\hspace{0.5cm}}
\caption{\label{PSI2_64} {\it {\footnotesize Evolution of the second normal stress coefficient plateau ($\Psi_2^{plateau}={N_2}/{(\dot{\gamma}\tau_0)^2}$) as a function of the shear rate. The scaling law observed experimentally~\cite{Schweizer2004},  $-\Psi_2 \sim \dot{\gamma}^{-1.6}$,  is shown for comparison.
Parameters~: $N_m=64$, $N_e=4$, $N_s=0.5$.}}}
\end{figure}

When it comes to the second normal stress difference $N_2$, we have not displayed the time evolution during shear flow, as it is much more noisy  than $N_1$ due to the 
low values of $N_2$. Rather we have measured the steady state value $N_2(\dot{\gamma})$ by averaging the instantaneous values of $N_2$ in a long time window such that 
the error bar in the determination of $N_2(\dot{\gamma})$ is a $20 \%$ typically. The resulting values of $\psi_2$ are shown in fig.~\ref{PSI2_64}. 
Again, at low shear rates $\psi_2$ is found to be a constant independent of $\dot{\gamma}$, with a $\psi_2/\psi_1$ ratio of order $-0.1$, typical of polymer systems. For stronger shear flow, $\psi_2$ decreases as 
$\psi_2 \propto \dot{\gamma}^{-\beta'}$ with $\beta' \simeq 1.5$ which is close to the exponent reported experimentally $\beta=1.6$\cite{Schweizer2004}.

\begin{figure}[H]
\centerline{\includegraphics*[width=8cm]{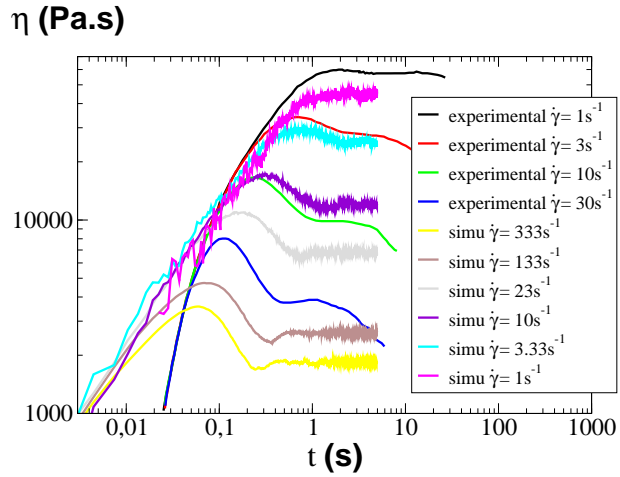}
\hspace{0.5cm}}
\caption{\label{viscosite_128_new} {\it {\footnotesize Comparison between the transient viscosities under steady shear flow obtained in the slip-link model and the experimental curves corresponding to polystyrene with a comparable number of entanglements per chain $Z=15$ (data taken from \cite{Schweizer2004}).  The two fitting parameters used here are $b=30.5$\AA\ and $\tau_0=3\times10^{-5}$s. The other slip-link parameters are $N_e=4$, $N_m=64$ and $N_s=0.5$.}}}
\end{figure}

\begin{figure}[H]
\centerline{\includegraphics*[width=8cm]{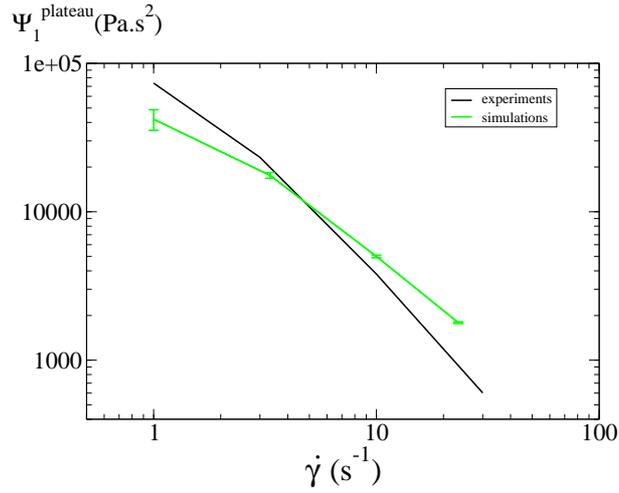}
\hspace{0.5cm}}
\caption{\label{psi1_plateau_comparaison} {\it {\footnotesize Comparison between the steady values of $\Psi_1$ and the experimental data of polystyrene having the same degree of entanglement (from \cite{Schweizer2004}). Same fitting and simulation parameters as in fig.\ref{viscosite_128_new}.}}}
\end{figure}

In Fig. (\ref{viscosite_128_new}), we compare the instantaneous viscosity $\eta(t)=\sigma_{xy}^{\rm Rouse+SL}(t)/\dot{\gamma}$ obtained from the Likhtman's model to experimental results for monodispersed polystyrene given in~\cite{Schweizer2004}. The experimental system is characterized by a number of entanglements per chain around $Z=15$ 
similar to our simulations ($Z=N_m/N_e=16$) and by chains made of $1920$ monomers which corresponds to a number of 
monomers per bead close to $30$. The two fitting parameters $b$ and $\tau_0$ used to scale the viscosity of the model $k_BT \tau_0/b^3$, have been tuned so as to minimize the absolute difference between the steady state viscosity obtained in our simulations and the  experimental data. This procedure leads to $b=30.5$\AA\ and $\tau_0=3\times 10^{-5}$s, which corresponds to roughly $30$ monomers per bead, and the correct order of magnitude for the corresponding Rouse time.  With this choice one sees from figure (\ref{viscosite_128_new}) that the family of simulation curves for the instantaneous viscosity as a function of time is consistent with the family of curves obtained from experiments at different shear rates. 
Although the instantaneous viscosity curves are reasonable, the experimental results in~\cite{Schweizer2004} are consistent with an effective  shear thinning exponent $0.86$, which is slightly higher than our simulation result $0.67$, so that the adjustment is not perfect. In Fig. (\ref{psi1_plateau_comparaison}), we display the evolution of $\psi_1^{\rm plateau}$ obtained using the same values  of fit parameters. The discrepancies between the simulation and the experimental data may be again attributed to the power law exponent $\Psi_1 \propto \dot{\gamma}^{-\kappa'}$ that is smaller in our simulations $\kappa' \simeq 1$ than in rheological measurements $\kappa' \simeq 1.5$.
% The experimental data are extracted from \cite{Schweizer2004}. 
\newline
In conclusion, the nonlinear flow properties of the model appear to be quite typical of what is experimentally observed in entangled polymer melts, although the effective shear-thinning exponents characterizing the normal stress coefficients are somewhat smaller than what is reported from rheological measurements. With this caveat, the slip-link model may be used to describe a ''generic'' polymer melt in complex situations, at a computational cost much lower than standard molecular dynamics simulations. We illustrate this point in the next section after extending the model to include spatial information and excluded volume interactions.

\section{Introducing excluded volume and space : a step towards modeling nanocomposites}
\begin{figure}[H]
\centerline{\includegraphics*[width=8cm]{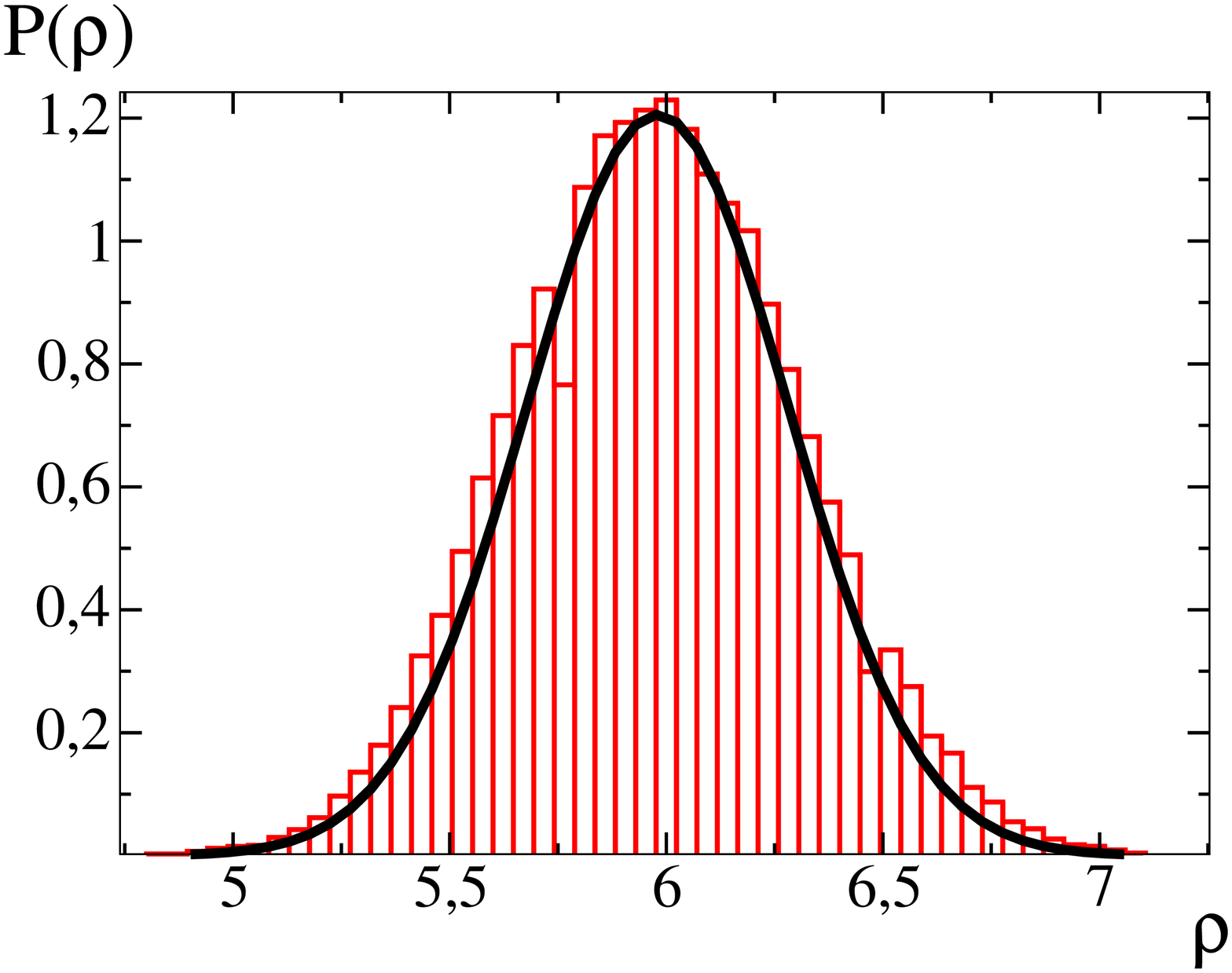}
\hspace{0.5cm}}
\caption{\label{histo} {\it {\footnotesize Monomer density distribution in cells of length $\delta=4.37$b for a polymer melt with excluded volume interaction $\kappa_0 N_m=50$, $\rho_0=6$b$^{-3}$, and with slip links ($N_m=64$, $N_e=4$, $N_s=0.5$). The black curve displays the theoretical distribution eq.~(\ref{distrib_density}). }}}
\end{figure}

So far, we have considered phantom polymer chains that can cross each other,  which is sufficient to describe homogeneous melts of homopolymers.
However, in most of the situations practically encountered,  polymer melts are not homogeneous.
This is the case for instance in nanocomposites, or in  thin films where the proximity of an interface affects the configurations of the 
polymer chains and the monomer density as well. In such situations, the polymer density results 
from the competition between the interaction between the monomers and  the surface, the entropy of the chains and the compressibility of the polymer melt.
To address such situations, it is necessary to introduce excluded volume interactions between segments in the slip link model. 
A relatively simple and computationally efficient way to account for these interactions is to consider a mean field version of
the excluded volume Hamiltonian, discretized on a lattice~\cite{Muller2008}:\\
\BEA
        \frac{\mathcal{H}_{hom}}{k_B\,T}=\frac{\kappa_0\,\delta^3}{2\rho_0} \, \sum_{\vec{c}} \, \left(\rho ( \vec{c} )-\rho_0 \right)^2
\label{hamiltonian_excluded_volume}
\EEA
where $\kappa_0$ is the dimensionless bulk modulus $\kappa_0 = 1/{k_BT \rho_0 \kappa_T}$ with  $\kappa_T=-\frac{1}{V}\left(\frac{\partial V}{\partial P}\right)_T$ being the compressibility, and $\rho_0$ is the mean segment density of the melt. The densities $\rho( \vec{c})$ are computed on a cubic lattice defined by the nodes $\vec c$, with $\delta^3$ being the volume of an elementary cell. The density $\rho(\vec c)$ is defined by the positions of the monomers in the vicinity of $\vec c$: 
\BEA
        \rho ( \vec{c} ) = \frac{1}{\delta^3} \sum_{n_c=1}^{N_p}\sum_{i=1}^{N_m} W(\vec r_i -\vec c) 
        \label{rho_lattice}
\EEA
with
\BEA
       W(\vec r_i -\vec c) = \prod_{\alpha=x,y,z}^{} \omega(r_{\alpha}-c_{\alpha})
\EEA
The weight function $W$ describes how each monomer contributes to the average density. Its values on the lattice of discrete sites $\vec c$ give the so called charge assignment functions  \cite{Deserno1998} of the particle located at point $\vec r$. They must, in particular, have the property that, 
\BEA
\forall \vec r, \quad \sum_{\vec c} W(\vec r -\vec c)=1
\EEA  
so that the lattice sum  of equation \ref{rho_lattice} gives the total number of particles.  In general, $W$ is chosen to be a short range function, that spreads the density associated with one particle over a few neighboring lattice sites. A convenient choice, due to Hockney and Eastwood (see ref.  \cite{Deserno1998}), is to take a function that spreads the particle over the $P$ neigboring nodes of the lattice. Assuming that the  lattice sites have integer coordinates (in units of the lattice spacing $\delta$) the charge assignment function of order $P$ is defined, in one dimension,  through
\BEA
W^{(P)}= \chi*W^{(P-1)}
\EEA  
where the$*$ denotes a convolution product, $\chi$ is the characteristic function of the interval $[-1/2,1/2]$ and $W^{(1)}=\chi$.   If we consider a particle with position (in units of the grid spacing) 
 $ 0<x<1$, clearly $W^{(1)}$ assigns the particle to the nearest lattice site with weight 1, $W^{(2)}$ assigns it to the nearest two sites 0 and 1 with weights 
\BEA
        &&\left\{
        \begin{array}{ccc}
        W_{0}^{(2)}(x) = 1-x \\
        W_{1}^{(2)}(x) = x \\
        \end{array}
        \right.
\EEA
We will also make use of the case where $P=4$, which gives charge assignment function on the 4 nearest nodes(-1,0, 1 and 2) 
of the form:
\BEA
\label{W4}
        &&\left\{
        \begin{array}{ccc}
        W_{-1}^{(4)}(x) &=& \frac{1}{6}(1-4x+4x^2-x^3) \\
	W_{0}^{(4)}(x) &=& \frac{1}{48}(32-48x^2+24x^3) \\
	W_{1}^{(4)}(x) &=& \frac{1}{6}(1+4x+4x^2-4x^3) \\
	W_{2}^{(4)}(x) &=& (x^3/6) \\
        \end{array}
        \right.
\EEA
The three dimensional assignment is achieved by using the product of the three assignment functions on each dimension, i.e. the particle density is spread over 8 nodes for $P=2$ and 64 nodes for $P=4$. We have simulated an ensemble of chains with slip links interacting through the Hamiltonian eq.~(\ref{hamiltonian_excluded_volume}).
Compared to the  previous simulations, each monomer $i$ feels the interaction force derived from the Hamiltonian eq.~\ref{hamiltonian_excluded_volume}:
\BEA
\vec{F}_{\rm hom} = - \frac{\kappa_0 \delta^3}{\rho_0} \, \sum_{\vec{c}_i} \left[ \left( \rho ( \vec{c}_i ) - \rho_0 \right)
 \vec{\nabla}_{\vec{r}_i} \left( \rho ( \vec{c}_i ) \right) \right]
 \label{f_hom}
\EEA
where the set $\vec c_i$ denotes the set of the $P^3$ node vectors nearest neighbors of the monomer $i$.
We have used the parameters $N_m=64$, $N_e=4$, $N_s=0.5$ for the slip links, and regarding the excluded volume interactions, 
$\kappa_0 N_m=50$, $\rho_0=5.98$ following~\cite{Muller2008}. We have simulated the dynamics of an ensemble of typically $1000$ chains 
and used a discretization length $\delta \sim 1.2 b$ for the calculation of the density fields. \newline 
After typically $1000$ time steps, the variance of the density fluctuations saturates, and we have checked that, under  these conditions, the 
Gaussian statistics of the chain is weakly affected by the excluded volume interaction.  
Figure \ref{histo} displays the monomer density distribution estimated by counting the number of monomers in large cells of length $\Delta \simeq 4.4$ b. 
Note that the discretization used for the estimate of the density here is not the same as the one used to calculate the density field 
in eq.~(\ref{hamiltonian_excluded_volume}). Figure~(\ref{histo}) shows that the actual monomer distribution is
well predicted by the thermodynamic expectation:
\BEA
P( \rho) = \frac{1}{\sqrt{2 \pi}} \exp \left(-\frac{(\rho-\rho_0)^2}{2 \; \sigma^2} \right)
\label{distrib_density}
\EEA
where $\sigma^2= \rho_0/(\Delta^3 \kappa_0)$ is the variance of density fluctuations at the scale $\Delta$ under consideration.
%This is truethe number of nodes $P$ per monomer used in the calculation of the density field.

\begin{figure}[H]
\centerline{\includegraphics*[width=8cm]{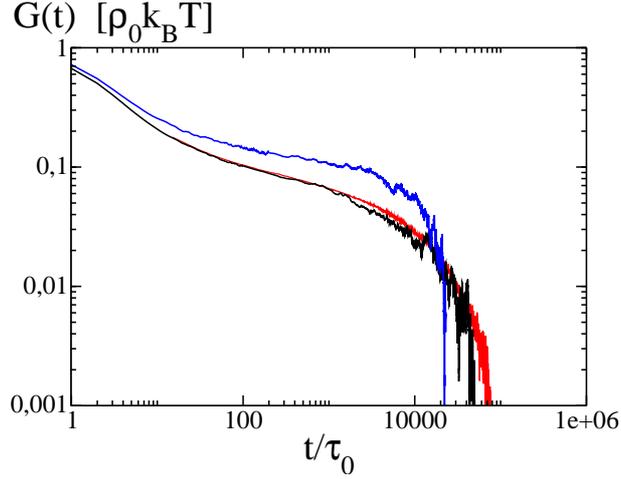}
\hspace{0.5cm}}
\caption{\label{G_kappa_varie} {\it {\footnotesize Stress relaxation modulus against time for a melt of ghost polymer chains with slip links (red curve) 
and for a melt of interacting chains. In this latter case, we have compared the result when a monomer contributes to the density of $P^3=8$ nodes (blue curve) and $P^3=48$ nodes (black curve). See text for further detail on the density discretization. The parameters are $\rho_0= 6$b$^{-3}$, $\kappa_0 N_m=50$. The other parameters retained are~:$N_m=64$, $N_e=4$, $N_s=0.5$.}}}
\end{figure}

\begin{figure}[H]
\centerline{\includegraphics*[width=8cm]{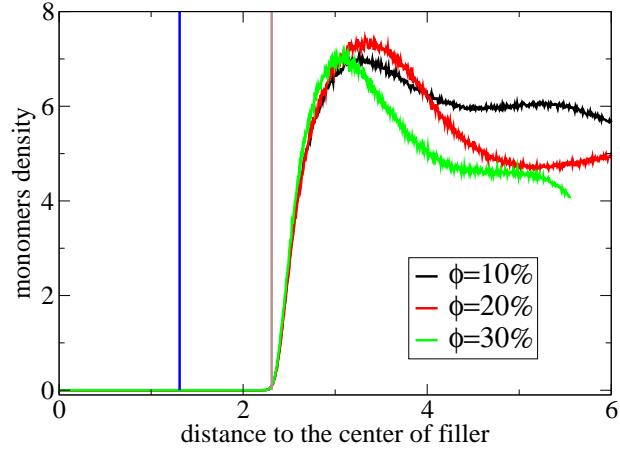}
\hspace{0.5cm}}
\caption{\label{densite} {\it {\footnotesize Monomer density as a function of the distance to the center of the filler for different filler volume fractions, $\phi=10$\%, $\phi=20$\% and $\phi=30$\%. In blue, we have represented the filler radius $\sigma_f$ while the brown line corresponds to the effective radius $\sigma_{eff}=\sigma_f+b$. We have considered $n_f=8$ fillers dispersed on a cubic lattice. The polymer parameters are $N_m=32$, $\kappa_0N_m=50$, $\rho_0=5.98$, $N_e=4$ and $N_s=0.5$.
}}}
\end{figure}

We have also assessed the dynamics of the polymer melt model with excluded volume interactions. To this end, we have compared the stress relaxation modulus with and without excluded volume interaction, 
The stress relaxation modulus $G(t)$ is computed using equilibrium simulations as explained in the previous sections (eq. (\ref{modulebis}) and (\ref{total_stress})).  Indeed the excluded volume interactions do not change the Green-Kubo expression of the shear relaxation modulus (eq. (\ref{modulebis})) since they generate only irrelevant pressure terms~\cite{DoiEdwards1986} and 
the total stress $\sigma_{\alpha\beta}^{T}$ is reduced to $\sigma_{\alpha\beta}^{R}+\sigma_{\alpha\beta}^{SL}$.
In presence of excluded volume interactions, it turned out that the resulting $G(t)$ depended on the discretization 
of the density field $\rho(\vec c)$, and in particular on the number of nodes $P$ where the density of a monomer is distributed.
This can be understood from the fact that with a discretization on only $P=2$ nodes, the force given by equation (\ref{f_hom}) is not a continuous function of space: when the particle crosses a cell, the nodes that contribute to the sum in equation (\ref{f_hom}) change, while their contribution to the force do not vanish, therefore introducing a discontinuity. On the other hand, for the $P=4$ scheme, the force induced by the lattice node that is farther away from the particle vanishes when this node ceases to be a 
neighbor. In general, the function $W^{(P)}(x)$  defined by the charge assignment of order $P$ is $P-2$ times differentiable, so that the minimum value of $P$ for which 
spurious force discontinuity can be, in principle, avoided is $P=3$.  As shown in figure~\ref{G_kappa_varie} , the choice  $P=4$ allows one to recover precisely the 
relaxation modulus  $G(t)$ of the chains without excluded volume, as expected from theoretical considerations on short range interactions \cite{DoiEdwards1986}. The 
Hamiltonian eq.~(\ref{hamiltonian_excluded_volume}), with the appropriate assignment of particles to the lattice, guarantees therefore a thermodynamically correct 
representation of excluded volume interactions without perturbing the dynamics of the chains. \newline
The last point to be discussed is the renewal rules for the slip links. Indeed, with the aim of introducing some spatial heterogeneities in the system, we must introduce a spatial constraint in the rules governing the destruction and rebirth of the slip-links.
To take into account the constraint release processes, we conserve the static binary correspondence between slip links. When a slip link passes through the end of its chain, 
it is instantaneously recreated at an extremity of a random chain $n_c$. However, the new chain $n_c$ is chosen so that its center of mass is at a maximal distance $R_g$ from  the  original slip link, where $R_g$ denotes the radius of gyration of the chains. The companion slip-link is also destroyed and instantaneously recreated at a random position in a random chain whose center of mass is again at a distance $R_g$ away from the center of mass of $n_c$. This spatial constraint seems natural since the diffusion of the center of mass of a 
chain must be small during the typical lifetime of a slip-link. Thus, an entanglement must be recreated in the vicinity of the destroyed one rather than anywhere in the system. To check wether these spatial rules do not lead to spurious effects, like {\em e.g.} an irreversible time-increasing concentration of coupled slip-links on the same chain, we have quantified the number of self-entanglements, {\it i.e} the number of pair of slip-links belonging to the same chain. This number has been found not to increase with time, and represents typically an amount of $5$ percents of the total number of entanglements, which is reasonable. \newline
\newline
We now apply this extension of Likhtman's model to the modeling of a filled entangled polymer melt. In  the following, we consider $n_f=8$ fillers distributed on a simple cubic lattice, with periodic boundary conditions. These fillers are modeled as fixed hard spheres with a radius $\sigma_f$. The filler-monomer interaction is taken to be repulsive:
 \BEA
\label{fill_force}
%   F_{fil}^{(i,n,j)}=\vec{\nabla}_{\vec{r}_i(n)} U_{fil}(\left\|\vec{r}_i\right\|,n,j)=\\
\vec{F}_{\rm fil}^{(i,n,j)}=F_{\rm fil}^{i,n,j}   \frac{\vec{r}_i(n)-\vec{r}_f^{\ j}}{\left\|\vec{r}_i(n)-\vec{r}_f^{\ j}\right\|} \\
\nonumber  {\rm with} \; F_{\rm fil}^{i,n,j}=\left\{
        \begin{array}{ccc}
	\frac{48k_BTb^{12}}{(\left\|\vec{r}_i(n)-\vec{r}_f^{\ j}\right\|\ -\sigma_{f})^{13}}& {\rm if} & \left\|\vec{r}_i(n)-\vec{r}_f^{\ j}\right\| > \sigma_{f}\\
	F_{\rm max} & {\rm if} & \left\|\vec{r}_i(n)-\vec{r}_f^{\ j}\right\| \le \sigma_{f} \\
        \end{array}
        \right.
\EEA  
where $\vec{F}_{\rm fil}(\left\|\vec{r}_i\right\|,n,j)$ is the force felt by the $i^{th}$ monomer of the $n^{th}$ chain due to the $j^{th}$ filler, $\vec{r}_i(n)$ represents the 
position of the $i$th monomer on the chain $n$ and $\vec{r}_f^{\ j}$ is the center of mass of the filler particle $j$. 
The modulus of the force $F_{\rm fill}$ is bounded by a maximal force $F_{\rm max}$ to avoid  very large forces, a situation encountered 
if a monomer is at a given time in the vicinity of a filler center of mass. We have taken typically $F_{\rm max}=100 k_B T/b$ for all the simulations. The additional repulsive force 
due to the presence of the fillers is simply added as an external force in the Langevin equations of motion of the monomers (Eq. (\ref{Langevin_mono})).
The steady monomer density profiles around a filler is represented in figure (\ref{densite}), for different values of the filler volume fraction. 
The volume fraction has been changed by tuning the volume of the system, keeping the number of fillers constant.
As a result of the filler repulsive interaction, the monomers are nearly totally excluded from an effective sphere of radius $\sigma_{\rm eff}=\sigma_f+b$ around the center of mass of the filler. The different density profiles beyond this exclusion zone result from the competition between the repulsive interaction between the monomers and the surface, the entropy of the chains and the compressibility of the polymer melt.

 The viscosity of the nanocomposite model can be  computed using equilibrium simulations and the Green-Kubo expression involving the integration of the stress stress correlation function:
\BEA
\label{modulebisbis}
        G(t)=\,\frac{V}{k_BT}\,\frac{1}{3}<\sum_{\alpha=1}^{2}\sum_{\beta>\alpha}^{3}\Bigg(\big(\sigma_{\alpha\beta}^{\rm Rouse}(t)+\sigma_{\alpha\beta}^{\rm fillers}(t)\big)\,\sigma_{\alpha\beta}^{\rm T}(0)\Bigg)>
\EEA
where $\sigma_{\alpha\beta}^{\rm fillers}$ is the instantaneous shear stress due to the filler-monomer interactions defined by
\BEA
\label{fill_stress}
\sigma_{\alpha\beta}^{\rm fillers}=-\frac{1}{V}\sum_{j=1}^{n_f}\sum_{n_c=1}^{N_p}\Big(\sum_{i=1}^{N_m}\alpha_i(n_c)F_{fil}^{(i,n_c,j)}(\beta)-\vec{r}_f^j(\alpha)\sum_{i=1}^{N_m}F_{fil}^{(i,n_c,j)}(\beta)\Big)
\EEA
where $n_f$, $N_p$ and $N_m$ are respectively the number of fillers, chains and monomers in the system of volume $V$. $\alpha_i(n_c)$ is the $\alpha$ coordinate vector of monomer $i$ of chain $n_c$. $\vec{r}_f^j(\alpha)$ is the $\alpha$ component of the position vector of filler $j$ and $F_{fil}^{(i,n_c,j)}(\beta)$ is the $\beta$ component of the force felt by monomer $i$ of chain $n_c$ due to filler $j$. 
Finally, 
\BEA
\label{total_stressbis}
\sigma_{\alpha\beta}^{\rm T}=\sigma_{\alpha\beta}^{\rm Rouse}+\sigma_{\alpha\beta}^{\rm SL}+\sigma_{\alpha\beta}^{\rm fillers}
\EEA 
denotes the total stress tensor, including the contribution of the Rouse forces, and the forces due to the slip-links and the fillers.  \\
Also, the filler volume fraction is defined here in terms of the effective radius $\sigma_{\rm eff}$, rather than using the bare value $\sigma_f$: the number of polymer chains in the system is:
   \BEA
\label{Np}
  N_p=\frac{4 n_f}{3 N_m}\pi\sigma_{\rm eff}^3\rho_0\Big(\frac{1}{\phi}-1\Big)
\EEA  
The key parameters and their values retained to model the polymer nanocomposite are summarized in table (\ref{table_parameters1}). 

\begin{figure}[H]
\centerline{\includegraphics*[width=10cm]{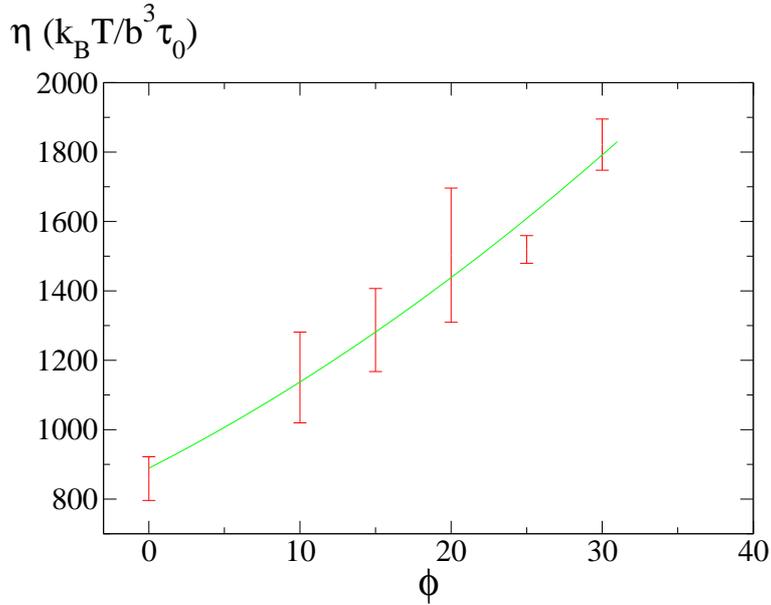}
\hspace{0.5cm}}
\caption{\label{reinforcement} {\it {\footnotesize Viscosity of the model nanocomposite as a function of the filler volume fraction $\phi$. The $n_f=8$ fillers are distributed on a cubic lattice. The different parameters used are summarized in Tab. (\ref{table_parameters1}).  We have also represented the fit obtained from the expression 
$\eta=\eta_0(1+\frac{5}{2}\phi+\beta\phi^2)$ where $\eta_0$ and $\beta$ are the two fitting parameters ($\eta_0=889\pm33$ $k_B T/b^3 \tau_0$ and $\beta=2.9\pm1.2$).
}}}
\end{figure}

In Fig. (\ref{reinforcement}), we show the evolution of the viscosity as a function of the filler volume fraction between $\phi=10$\% and $\phi=30$\%. As shown in this figure, the viscosity is well described by   the expression $\eta=\eta_0(1+\frac{5}{2}\phi+\beta\phi^2)$, classically used to describe the viscosity of dense suspensions. The fitting parameters are the viscosity $\eta_0$ and the coefficient $\beta$, which take the values  $\eta_0=889\pm33 k_BT/b^3 \tau_0$ and $\beta=2.9\pm1.2$. This Einstein like increase of the viscosity is maybe not surprising for a well dispersed filler suspension, in the absence of additional entanglements between the fillers and the polymer matrix. It  shows however that slip links models \`a la Likhtman may be extended to model the rheology of polymer nanocomposites at a relatively low cost.   Investigation of  the  dispersion state of filler particles, or of  additional entanglements with polymer chains grafted on the filler is possible and will be reported in further publications.

\begin{table}[H]

\begin{center}
\begin{tabular}{|c|c|}
\hline
%\bf{processus} & \bf{taux de transition} \\
%\hline
temperature & $k_B\,T=1$ \\ \hline
monomer size & $b=1$ \\ \hline
mean density of the polymer melt & $\rho_0=5.98$ \\ \hline
 dimensionless bulk modulus & $\kappa_0=50/N_m$  \\ \hline
filler volume fraction & $\phi\in \left[10\%\ ;\ 30\%\right] $ \\ \hline
effectif radius of fillers &  $\sigma_{eff}=R_g+b\approx 2.31$ \\ \hline
friction coefficient of the entropic springs & $\xi=1$ \\ \hline
friction coefficient of the slip-links & $\xi_S=0.1\,\xi$ \\ \hline
number of fillers & $n_f=8$ \\ \hline
number of monomers per chain & $N_m=32$ \\ \hline
number of Kuhn's segments between slip-links & $N_e=4$ \\ \hline
number of slip-links per chain & $Z=\frac{N_m}{N_e}=8$ \\ \hline
stiffness of the slip-links & $\frac{3k_BT}{N_sb^2}$ with $N_s=0.5$\\ \hline
characteristic time & $\tau_0=\frac{\xi\,b^2}{3\,\pi^2\,k_B\,T}$ \\ \hline
\end{tabular}
\end{center}
\label{table_parameters1}
\caption{Main parameters that define the slip link model applied to a nanocomposite. We have also indicated the values of the  parameters used in this work.}
\end{table}

\newpage

\section{Summary}

The slip link model studied in this manuscript has a number of attractive features that make it well suited for investigating the mechanical and rheological properties of 
complex polymer systems, at a level of coarse graining and over time scales that are far greater than those usually studied in molecular dynamics simulations. Specifically,
the linear rheology properties are close to those predicted by the reptation model of Doi and Edwards, but a greater flexibility is possible through the independent variation
 of the various model parameters. This was already demonstrated in the work of Likhtman, who showed the ability of the model to reproduce the linear rheology and spin echo 
data on a number of different polymer melts. The nonlinear rheology properties appear to be quite 'typical' of what is observed in entangled polymer melts.  
It also appears that these properties are maintained when introducing excluded volume (or more generally, specific interactions between different monomers)  in a mean field 
manner, in the spirit of what has been achieved at a smaller level of coarse graining~\cite{Detcheverry2009}.
The flexibility of slip-links models paves the ways to model nanocomposites, which display a hierarchy of length and times scales which makes the direct use of molecular dynamics simulations prohibitive.   Here, we have concentrated on an idealized situation   where the fillers are well dispersed, 
with a simple hardcore interaction between the fillers and the polymer matrix. Addressing real situations where the fillers are poorly dispersed and partially aggregated is clearly possible within the same framework.  Also, slip-links models offer the opportunity to tune the polymer/filler interaction, and introduce glass transition effects through the monomer friction coefficient. 
This will be the object of future investigations.
% Finally, we show that, using the excluded volume interactions, 
%we can extend the slip-link model developed by Likhtman to describe a particular nanocomposite system namely with bare fillers distributed on a cubic lattice. This study which is a first approach shows that the evolution of the viscosity follows an Einstein-like law. It allows to conclude that slip links models \`a la Likhtman may be extended to model the rheology of polymer nanocomposite at a relatively low cost. The goal, in the long term, would be 
%to use this extension to investigate the effect of fillers dispersion state and that of grafted chains (on fillers) on the %reinforcement.  

\section{Comment}
During the submission process, we became aware of two very recent articles~\cite{Chappa2012,Ramirez-Hernandez2013}, where the non-linear rheology of a  similar slip-link model (with a slightly different implementation)  has been investigated. The shear-thinning exponents for the viscosity and normal stress differences have been found to be close to our present findings~\cite{private_communication}, which indicates that they are quite independent from the specific scheme used for the slip link implementation.

\bf{Acknowledgements:}  JLB is supported by the Institut Universitaire de France ; we thank Juan de Pablo for sharing with us some preliminary results on the nonlinear rheology of a related slip link models.
%and for this reason we hypothesize that the agreement between their simulations and experimental data would be as good as our figure~\ref{viscosite_128_new}.  

%%%%%%%%%%%%%%%%%%%%%%%%%%%%%%%%%%%%  COURBES %%%%%%%%%%%%%%%%%%%%%%%%%%%%%%%%%%%%%%%%%%%%%%%%%%%%%%

%\section{Influence of model parameters on the relaxation modulus}
\newpage

%\section{Nonlinear rheological behavior}



%\section{Introducing excluded volume and spatial information}


${}$\\
%\bibliography{SL_elian}
%Merlin.mbs v4.21 2009-07-09.
%

\end{document}